\documentclass[lettersize,journal]{IEEEtran}
\usepackage{amsmath,amsfonts}
\usepackage{algorithmic}
\usepackage{algorithm}
\usepackage{array}
\usepackage[caption=false,font=normalsize,labelfont=sf,textfont=sf]{subfig}
\usepackage{textcomp}
\usepackage{stfloats}
\usepackage{url}
\usepackage{verbatim}
\usepackage{graphicx}
\def\BibTeX{{\rm B\kern-.05em{\sc i\kern-.025em b}\kern-.08em
    T\kern-.1667em\lower.7ex\hbox{E}\kern-.125emX}}
\usepackage{balance}

\usepackage{cite}

\usepackage{booktabs}
\usepackage{multirow}
\usepackage{adjustbox}
\usepackage{graphicx}
\usepackage{lipsum}
\usepackage{verbatim}
\usepackage{url}
\usepackage{amssymb}

\begin{document}

\title{Utilizing Neural Transducers for Two-Stage Text-to-Speech via Semantic Token Prediction}

\author{Minchan Kim~\IEEEmembership{{Student Member,~IEEE}}, Myeonghun Jeong~\IEEEmembership{{Student Member,~IEEE}}, Byoung Jin Choi~\IEEEmembership{{Student Member,~IEEE}}, Semin Kim, Joun Yeop Lee, and Nam Soo Kim~\IEEEmembership{Senior Member,~IEEE}

\thanks{Minchan Kim, Myeonghun Jeong, Byoung Jin Choi, Semin Kim, and Nam Soo Kim are with the Institute of New Media and Communications, Department of Electrical and Computer Engineering, Seoul National University, Seoul
08826, Republic of Korea (e-mail: mckim@hi.snu.ac.kr; mhjeong@hi.snu.ac.kr; bjchoi@hi.snu.ac.kr; smkim21@hi.snu.ac.kr; nkim@snu.ac.kr)
}
\thanks{Joun Yeop Lee is with Samsung Research, Seoul, 
06765, Republic of Korea (e-mail:  jounyeop.lee@samsung.com)
}}


\markboth{Journal of \LaTeX\ Class Files,~Vol.~14, No.~8, August~2021}%
{Shell \MakeLowercase{\textit{et al.}}: A Sample Article Using IEEEtran.cls for IEEE Journals}

\maketitle
\begin{abstract}
We propose a novel text-to-speech (TTS) framework centered around a neural transducer. Our approach divides the whole TTS pipeline into semantic-level sequence-to-sequence (seq2seq) modeling and fine-grained acoustic modeling stages, utilizing discrete semantic tokens obtained from wav2vec2.0 embeddings. For a robust and efficient alignment modeling, we employ a neural transducer named token transducer for the semantic token prediction, benefiting from its hard monotonic alignment constraints. Subsequently, a non-autoregressive (NAR) speech generator efficiently synthesizes waveforms from these semantic tokens. Additionally, a reference speech controls temporal dynamics and acoustic conditions at each stage. This decoupled framework reduces the training complexity of TTS while allowing each stage to focus on semantic and acoustic modeling. Our experimental results on zero-shot adaptive TTS demonstrate that our model surpasses the baseline in terms of speech quality and speaker similarity, both objectively and subjectively. We also delve into the inference speed and prosody control capabilities of our approach, highlighting the potential of neural transducers in TTS frameworks.

\end{abstract}

\begin{IEEEkeywords}
speech synthesis, neural transducer, zero-shot adaptive TTS, speech representation.
\end{IEEEkeywords}

\section{Introduction}
\IEEEPARstart{N}{eural} text-to-speech (TTS) systems have made significant progress in various aspects over the last few years. When designing a neural TTS model, alignment modeling is usually treated as one of the most critical components. Alignment modeling in TTS addresses the challenges of handling input-output length mismatches and establishing a proper alignment between text and speech. Since text units are expected to be monotonically aligned with speech frames, effectively harnessing this strong inductive bias is essential for achieving robust and high-quality TTS performance. Alignment modeling also plays a crucial role in enhancing phonetic intelligibility, as it governs the representation of linguistic information in speech. In recent TTS models, two primary approaches have been employed: attention-based autoregressive (AR) method and duration-based non-autoregressive (NAR) method. Attention-based AR models, like those proposed in \cite{shen2018natural, li2019neural, kharitonov2023speak}, operate using an AR model that predicts speech in a frame-by-frame manner and utilizes an attention mechanism to establish alignment. In contrast, duration-based NAR models, such as \cite{ren2020fastspeech, kim2021conditional, popov2021grad}, require phoneme-wise duration to regulate speech frame length and generate frames in parallel, necessitating external alignment search algorithms~\cite{mcauliffe2017montreal, kim2020glow} and explicit duration predictors to obtain durations.

Another critical challenge in TTS is addressing the diversity of speech that arises from the one-to-many relation between a given text and its spoken representation. The variation in speech can be attributed to paralinguistic characteristics such as speaker identity, prosody, and environmental conditions. To capture this variability, neural TTS models should be able to express these diverse characteristics. Traditionally, deep generative models such as variational autoencoders~(VAE)~\cite{kim2021conditional,xie2021disentangled}, normalizing flows~\cite{kim2020glow,kim2021conditional,miao2020flow}, and diffusion models~\cite{popov2021grad,jeong2021diff,shen2023naturalspeech} have been used to represent speech diversity. Additionally, auxiliary conditioning information is leveraged to control this variability, for example, by using a reference speech to represent the target speaker for zero-shot adaptive TTS. However, the presence of acoustic diversity complicates the data distribution, making TTS model training unstable and weakening alignment modeling, which is essential for maintaining a high level of intelligibility, a non-negotiable factor in TTS. Consequently, many existing TTS research endeavors focus on meticulously curated TTS datasets, such as \cite{ito2017lj, veaux2017cstr}, without extending their scope to encompass in-the-wild datasets, despite the potential for capturing more expressive variations.

In this paper, we present a novel TTS framework designed to address the aforementioned challenges. Our framework adopts a two-stage approach that leverages semantic tokens obtained from wav2vec2.0 embeddings~\cite{baevski2020wav2vec}. These semantic tokens contain rich linguistic information and are properly aligned with the speech frames. In the first stage, this framework converts text sequences into semantic token sequences, emphasizing linguistic and alignment modeling. In contrast, the second stage generates speech waveforms from the semantic token sequence, focusing on non-linguistic acoustic attributes.

In the first stage, we apply a neural transducer~\cite{graves2012sequence, zhang2020transformer}, referred to as the token transducer. The neural transducer, e.g., RNN-T~\cite{graves2012sequence}, is a sequence-to-sequence~(seq2seq) model with a monotonic alignment constraint. It searches for monotonic paths between the input and output sequences while computing the conditional likelihood using alignment lattices and dynamic programming. Though neural transducers are widely adopted in automatic speech recognition~(ASR), where the alignment assumption is congruent to TTS, their application to TTS is limited due to the continuous output space of TTS. Our approach bridges this gap by utilizing neural transducers to generate semantic tokens, which have a discrete output space and disentangled linguistic information. The proposed token transducer effectively captures the linguistic information of the semantic tokens, focusing on the alignment search.

The second stage is responsible for generating high-fidelity speech waveforms from the semantic token sequence. Since the semantic tokens are already aligned with speech frames, we employ a duration-based non-autoregressive (NAR) architecture to achieve fast inference speed. Our speech generator follows a VITS-based architecture~\cite{kim2021conditional} to ensure high-quality speech generation and incorporates reference speech as a conditioning input to control non-linguistic speech diversity. This design enables our proposed model to operate in a zero-shot adaptive TTS manner.

The contributions of our paper can be summarized as follows\footnote{This work is an extended research of our previous work~\cite{kim2023transduce}. We enhance the elucidation of the overall concepts within the proposed method and conduct a more comprehensive analysis across various aspects.}:
\begin{itemize}
\item We propose a novel TTS framework that incorporates two stages, utilizing semantic tokens from wav2vec2.0 embeddings to streamline training complexity, and enhance the robustness and efficiency of both stages by separating semantic and acoustic modeling.
\item We introduce a token transducer for semantic token prediction, capitalizing on the robust monotonic alignment constraint of the neural transducer. Additionally, we propose a VITS-based speech generator for high-fidelity and rapid inference.
\item Our experimental results in zero-shot adaptive TTS demonstrate that the proposed framework outperforms the baseline models in terms of intelligibility, naturalness, and zero-shot speaker adaptation.
\item We further explore the inference speed and prosody controllability of the proposed token transducer, highlighting the advantages of employing a neural transducer in the TTS framework.
\end{itemize}

The remainder of this paper is organized as follows: In Section II, we provide background knowledge on our work. In Section III, we describe the proposed method, which includes the token transducer and the speech generator. Section IV details the experimental settings, and Section V presents the results. Finally, we conclude our work in Section VI. We also upload the audio samples on our demo page: https://gannnn123.github.io/token-transducer

\section{Backgrounds}

\label{sec:format}

\subsection{Self-Supervised Representations for TTS}
In recent years, significant strides in the neural TTS have been driven by the incorporation of self-supervised learning~(SSL) based speech representations \cite{baevski2020wav2vec, chung2021w2v, hsu2021hubert}. These SSL models are trained to extract contextualized representations from speech waveforms through optimizing some self-supervision objectives, such as masked prediction of hidden units \cite{hsu2021hubert}. These representations are highly disentangled to contain phonetic and linguistic information and have been widely adopted in language-related applications, including speech recognition and speech translation. In the field of TTS, several works \cite{kim2022transfer, lee2022hierspeech, choi2022nansy, siuzdak2022wavthruvec, kharitonov2023speak} have demonstrated the potential of SSL representations as an intermediary embeddings, bridging the realms of text and speech. The disentangled nature of SSL representations is helpful in segregating linguistic and acoustic modeling, simplifying the complexity of TTS training. Furthermore, since SSL representations can be extracted from unlabeled speech corpora, they empower the TTS models to be trained on much larger datasets, consequently augmenting the quality and diversity of synthesized speech.

For instance, \cite{kim2022transfer} proposes a transfer learning framework for TTS using wav2vec2.0 embeddings for pre-training. This transfer learning framework pre-trains a model using wav2vec2.0 embeddings instead of text input, followed by fine-tuning with text input, proving its effectiveness, especially in cases with limited labeled datasets. Additionally, \cite{choi2022nansy, siuzdak2022wavthruvec, kharitonov2023speak} divide TTS systems into two distinct stages, employing SSL representations as an interface that interconnects these stages. These models train each stage independently and then cascade them for speech generation. Among these works, the work most relevant to ours is SPEAR-TTS \cite{kharitonov2023speak}. SPEAR-TTS leverages semantic tokens extracted from w2v-BERT \cite{chung2021w2v} and comprises two seq2seq stages, text-to-semantic token and semantic token-to-speech, respectively denoted as ``reading" and ``speaking" procedures. It is worth noting, however, that SPEAR-TTS relies on an attention-based encoder-decoder transformer for the ``reading" and a prompt-based codec language model for the ``speaking", whereas our methodology places greater emphasis on monotonic alignment and efficient speech generation.

\subsection{Alignment Modeling in Neural TTS}
Alignment modeling deals with the task of establishing a correct alignment between text and speech. There are two primary approaches to alignment modeling: the attention-based autoregressive (AR) method and the duration-based non-autoregressive (NAR) method.

\subsubsection{\bf{Attention-based AR method}}
Attention mechanisms~\cite{bahdanau2014neural, soydaner2022attention} have found widespread application in the domain of seq2seq models, especially when the alignment between input and output is not explicitly defined. Attention-based AR models~\cite{shen2018natural, li2019neural, kharitonov2023speak} reframe seq2seq models as AR processes as follows:
\begin{equation} \label{eq1}
p(\textbf{y}|\textbf{x}) = p(y_{0}|\textbf{x}) \prod_{t=1}^{T} p(y_{t}|y_{<t},\textbf{x}),
\end{equation}
where $\textbf{x}=\{x_u\}^{U}_{u=1}$ denotes the input sequence and $\textbf{y}=\{y_t\}^{T}_{t=0}$ represents the output sequence. Here, $U$ and $T$ are the lengths of input and output sequences, respectively, and $y_0$ corresponds to the start of the sequence, i.e., $\langle SOS\rangle$ token. In this approach, speech frames are generated sequentially to circumvent the length mismatch between speech and text. During each prediction of $y_t$, the attention mechanism computes alignment scores for $y_t$ using Key~(K), Query~(Q), and Value~(V) matrices. The query Q is derived from $y_t$, while the K and V are extracted from $\textbf{x}$. The attention scores establish a soft alignment between $\textbf{x}$ and $\textbf{y}$.

Although the attention mechanism can discover alignment without any explicit guidance, it does not guarantee a monotonic alignment and may involve redundant calculations for Q-K matching. Consequently, attention-based AR models usually face potential misalignment issues \cite{valentini2021detection}, such as word skipping, repetitions, and early terminations. Previous works~\cite{liang2020enhancing, tachibana2018efficiently, georgiou2023regotron} have been proposed to enhance monotonicity in attention mechanisms. However, these approaches only assist in maintaining monotonicity rather than enforcing an efficient and strict monotonic constraint.

\subsubsection{\bf{Duration-based NAR method}}
The duration-based NAR method involves explicitly determining the duration for each text unit. Within this approach, text embeddings are duplicated according to their pre-determined durations to align with speech frames. For training, ground truth durations are obtained from the pairs of text and speech using external monotonic alignment algorithms~\cite{mcauliffe2017montreal, kim2020glow}. During inference, when ground truth durations are inaccessible, an explicit duration predictor infers durations from text representations instead. Once the alignment between text representations and speech frames is established, various NAR decoder structures can be utilized to convert the length-extended text embeddings into speech frames via parallel computation. Using explicit durations, these models mitigate alignment issues, rendering them more robust and stable during both training and inference. Moreover, the use of NAR decoder structures results in faster inference speed compared to AR methods.

Nevertheless, these methods still have limitations. First, the disparity in duration extraction between training and inference can degrade sample quality since durations from the duration predictor are not experienced during training. When the distribution of duration becomes complex and exhibits one-to-many mappings due to speech diversity, developing an optimal duration predictor becomes more challenging, exacerbating this mismatch. Second, duration-related information, such as speech rate or prosody, cannot be jointly modeled with other attributes due to the hierarchical nature of generating duration and speech frames. This hierarchy may impede the expression of fine-grained speech details.

\subsection{Neural Transducer}
\subsubsection{\bf{Formulation}}
Neural transducers~\cite{graves2012sequence} establish a seq2seq model with a monotonic alignment constraint between $\textbf{x}$ and $\textbf{y}$, where $\textbf{x}$ and $\textbf{y}$ are defined as in (1). As illustrated in Fig.~1(a), the neural transducer constructs a lattice $\{\alpha(u,t)|{1\le u\le {U}}, {0\le t\le {T}}\}$ to compute the probability of $P(\textbf{y}|\textbf{x})$. Each node $\alpha(u,t)$ represents $P(y_{0:t}|x_{1:u})$, indicating the probability of emitting the initial output sequence $y_{0:t}$ corresponding to the input sequence $x_{1:u}$. For each node, the neural transducer calculates the emission probability $P(y_{t+1}|u,t)$ and transition probability $P(\varnothing|u,t)$. The $\langle blank\rangle$ token $\varnothing$ signifies the transition to the next input frame, indicating the horizontal arrows in Fig.~1(a). Thus, the output of the transducer is in the space of $\mathcal{V}\cup \{\varnothing\}$, where $\mathcal{V}$ represents the actual output space. The training objective of the neural transducer is defined as follows:
\begin{equation} \label{eq1}
\begin{split}
\mathcal{L}_{tran} &= -\log{P(\textbf{y}|\textbf{x})} \\
&= -\log{\sum_{a \in \mathcal{F}^{-1}(Y)}{P(a|\textbf{x})}}.
\end{split}
\end{equation}
In (2), $a$ represents a possible monotonic path, and $\mathcal{F}^{-1}$ refers to the inverse function of the removal function $\mathcal{F}$. For example, in Fig.~1(a), an example alignment between $\textbf{x}$ and $\textbf{y}$ is denoted as $\hat{a}=(y_{1},y_{2},\varnothing,\varnothing,y_{3},y_{4},\varnothing,y_{5},\varnothing)$, and $\mathcal{F}(\hat{a})=(y_{1},y_{2},y_{3},y_{4},y_{5})$. Consequently, $P(\textbf{y}|\textbf{x})$ is derived as the marginalized likelihood of all possible monotonic paths, satisfying $\mathcal{L}_{trans}=\alpha(U,T)P(\varnothing|U,T)$. The lattice nodes can be dynamically calculated using the following forward algorithm:
\begin{equation} \label{eq:loss2}
\begin{split}
  \alpha(u,t) = \alpha(u-1,t)P(\varnothing|u-1,t) \\+ \alpha(u,t-1)P(y_{t}|u,t-1).
\end{split}
\end{equation}
For inference, the neural transducer generates the output sequence autoregressively. The aforementioned formulations allow the neural transducer to efficiently search for monotonic alignments and generate sequences based on the monotonic alignment constraint.
\begin{figure}[]
 \centering
 \includegraphics[width=0.9\columnwidth]{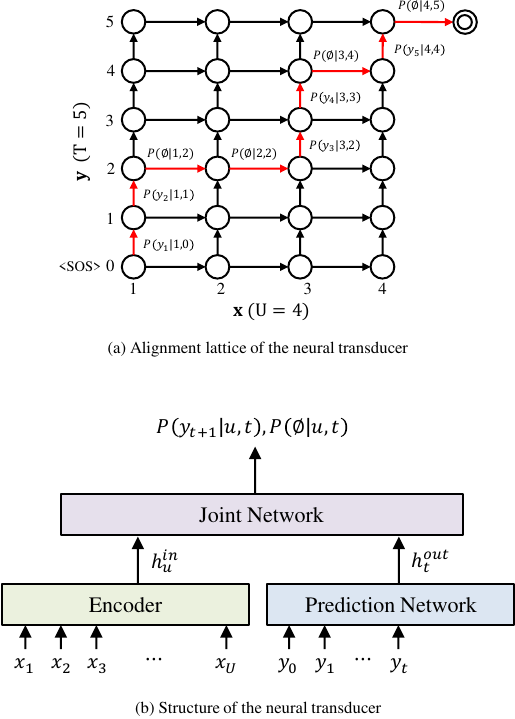}
 \caption{Description of a neural transducer. The red path in (a) represents an example of a monotonic path.}
 \label{fig:t-SNE}
 \end{figure}
 
\subsubsection{\bf{Architecture}}
As shown in Fig~1.(b), a neural transducer comprieses three integral modules: an encoder, a prediction network, and a joint network. The encoder takes an input sequence $\textbf{x}$ and yields a hidden embedding sequence $\textbf{h}^{in}=\{h_u^{in}\}^{U}_{u=1}$. Meanwhile, the prediction network operates in an autoregressive fashion, taking the previous output frames and returning the hidden output embedding sequence $\textbf{h}^{out}=\{h_t^{out}\}^{T}_{t=0}$. Consequently, the prediction network can be implemented by causal sequence models, such as unidirectional LSTM or causally masked transformers. The joint network processes $\textbf{h}^{in}$ and $\textbf{h}^{out}$ and outputs the probabilities for the elements in $\mathcal{V}\cup \{\varnothing\}$ for each node. Specifically, the joint network takes $h^{in}_u, h^{out}_t$ as input, and returns $P(y_{t+1}|u,t)$ and $P(\varnothing\ |u,t)$. Notably, the joint network computes the probabilities for individual embedding vectors, rather than the entire sequence, permitting the utilization of a simple projection layer or fully connected layers in its design.

\subsubsection{\bf{Application}}
Neural transducers serve as powerful tools across a wide range of seq2seq tasks that adhere to the monotonic alignment assumption. While these tasks encompass a wide array of applications, neural transducers have predominantly unfurled their potential in automatic speech recognition~(ASR)~\cite{prabhavalkar2017comparison, zhang2020transformer, dalmia2021transformer}. Their specific advantages come to the fore in streaming ASR scenarios~\cite{he2019streaming, sklyar2021streaming, sklyar2022multi}, primarily owing to the intrinsic monotonic constraint they embrace. However, while TTS shares several commonalities with ASR, adopting neural transducers for TTS introduces several challenges.
 
Traditionally, the output space of TTS, often dwelling in the continuous domain, such as mel-spectrograms, pose a significant challenge when it comes to the concurrent modeling of both continuous emission probabilities and discrete transition probabilities. Prior studies, such as SSNT-TTS \cite{yasuda19_ssw, yasuda2020effect} and Speech-T \cite{chen2021speech}, have explored the application of neural transducers in TTS. SSNT-TTS aimed to directly model the joint probability of transitions and mel-spectrogram emissions, while Speech-T introduced the lazy-forward algorithm to separate transition and emission modeling effectively. However, owing to the intrinsic one-to-many nature of TTS, these regression-based prediction methods encounter difficulties in capturing diversity. Consequently, they were applied to single-speaker TTS applications with restricted variability rather than proving their competitivity in more diverse and complex scenarios.

\section{Proposed Method}
\label{sec:pagestyle}
In this section, we propose a novel TTS model, which consists of two components: a token transducer and a speech generator. To begin with, we define the semantic token as the index of the $k$-means clustering on wav2vec2.0 embeddings. The notable characteristics of the semantic token are that 1) the semantic token sequence is aligned with speech frames, and 2) the semantic token contains disentangled linguistic information and is robust to fine-grained acoustic variations. Leveraging these properties, we partition the TTS task into two distinctive subtasks: text-to-token and token-to-speech. In the following subsections of III-A and III-B, we provide detailed descriptions of each component.

\subsection{Token Transducer}
We employ a neural transducer, named the token transducer, for semantic token prediction. The token transducer plays a primary role in alignment modeling and encoding the semantic information of the text sequence. Targeting the semantic tokens helps the neural transducer overcome the conventional limitations of applying a neural transducer for TTS. Quantization via $k$-means clustering renders the output space discrete, eliminating the necessity for complex modeling for the continuous output space. Furthermore, using semantic tokens reduces data complexity attributed to fine-grained speech conditions, allowing the neural transducer to focus on alignment and semantic information.

\subsubsection{\bf{Formulation}}
We follow the formulation of the neural transducer described in Section II-C. Here, the input sequence $\textbf{x}$ corresponds to a text, while the output sequence $\textbf{y}$ represents the semantic token sequence. One notable departure from the basic neural transducer is the additional conditioning factor, the reference speech~$\textbf{s}_{ref}$. In contrast to the text sequence, semantic tokens encompass more extensive information concerning duration and prosody variations. Thus, we introduce the reference speech as a conditioning factor for the token transducer to control the supplementary information. Consequently, we reformulate the modeling objective as $P(\textbf{y}|\textbf{x},\textbf{s}_{ref})$, thereby modifying the training objective as follows:
\begin{equation} \label{eq:loss1}
  \mathcal{L}_{tran'} = -\log{\sum_{a \in \mathcal{F}^{-1}(\textbf{y})}{P(a|\textbf{x}, \textbf{s}_{ref})}}.
\end{equation}

\subsubsection{\bf{Architecture}}
A prototypical architecture of the proposed token transducer is illustrated in Fig.~2. The token transducer consists of four components: a text encoder, a prediction network, a reference encoder, and a joint network. For the text encoder, we deploy conformer blocks~\cite{gulati2020conformer}. For the prediction network, we explore and compare two structural choices: unidirectional LSTM and masked conformer. Given the autoregressive nature of the prediction network, it serves as the most critical bottleneck impacting inference speed. The LSTM-based prediction network offers fast generation speed thanks to its efficient recursion, whereas the conformer-based prediction network requires more computation resources with $O(T^2)$ complexity, offering a higher model capacity in return. The reference encoder takes a reference speech $\textbf{s}_{ref}$ and yields a reference embedding vector $h^{ref}$. We employ the architecture of ECAPA-TDNN~\cite{desplanques2020ecapa} for the reference encoder, which is jointly optimized based on the transducer objective. Finally, the joint network takes the summation of the outputs of the text encoder and the prediction network, i.e., $h_u^{in}+h_t^{out}$, and returns the output probabilities, $P(y_{t+1}|u,t)$ and $P(\varnothing|u,t)$. The joint network is composed of residual blocks of linear layers and conditional layer normalization layers. The reference embedding $h^{ref}$ is projected to the scale factors of the conditional layer normalization layers. We emphasize that we are more focused on the generative capacity of the neural transducer compared to the conventional transducer for ASR, given a richer information encoded within the semantic tokens. As a result, we adopt a relatively more expressive architecture for the prediction network and the joint network, where a stateless prediction network and a single projection joint network demonstrate comparable capacity for ASR performance~\cite{ghodsi2020rnn}.

\subsubsection{\bf{Pruned training method}}
Training a neural transducer is notorious for its high memory consumption, with the primary memory-consuming factor being the construction of the lattice by the joint network at a scale of $O(UT)$ memory complexity. To enlarge the capacity of the joint network, we apply a pruned training method~\cite{kuang2022pruned}. In this method, we identify a candidate region suitable for plausible alignment and exclude unnecessary nodes. Subsequently, the joint network performs computations exclusively on the pruned nodes, limited to $S$ nodes for each text frame, where $S$ represents a pruning constant. Pruned training effectively reduces the training memory complexity from $O(UT)$ to a much more manageable $O(U)$. Finally, we set the training objective to be $\mathcal{L}_{tran''} = \alpha_{1}\mathcal{L}_{simple} + \alpha_{2}\mathcal{L}_{tran'}$, where $\mathcal{L}_{simple}$ is the loss for searching the pruned range, and $\mathcal{L}_{tran'}$ is calculated in the pruned region. Also, $\{\alpha_{1},\alpha_{2}\}$ serve as scale factors. For more comprehensive details on the pruned training method, we encourage interested readers to refer to \cite{kuang2022pruned}.

\begin{figure}[]
 \centering
 \includegraphics[width=0.9\columnwidth]{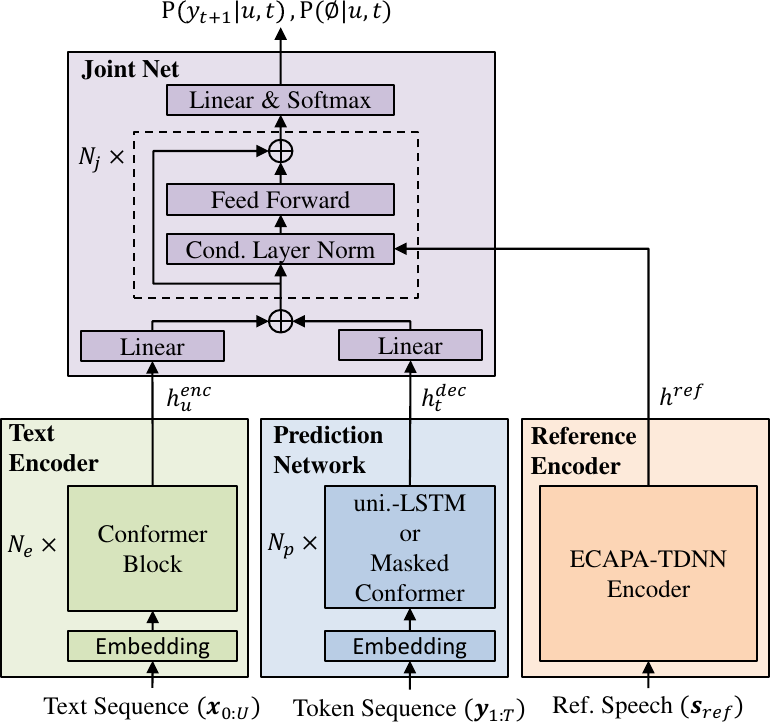}
 \caption{Architecture of the proposed token transducer.}
 \label{fig:t-SNE}
 \end{figure}

\subsection{Speech Generator}
The speech generator converts the semantic token sequence into the corresponding speech waveform, focusing on generating high-fidelity audio while controlling acoustic conditions using a reference speech. Since the semantic tokens are aligned with speech frames, there are no issues pertaining to length variability or alignment in speech generation. For that reason, we design the speech generator with an NAR architecture due to its fast and efficient generation ability. We leverage the VITS~\cite{kim2021conditional} architecture as the foundation for our speech generator.

\subsubsection{\bf{VITS overview}}
VITS is an end-to-end TTS model based on an NAR structure and a conditional variational autoencoder~(CVAE) formulation. VITS consists of a posterior encoder, a prior encoder, and a decoder. It manages the conditional likelihood of the speech waveform~$\textbf{w}$ by maximizing the evidence lower bound~(ELBO) formulated as follows:
\begin{equation} \label{eq:loss1}
  \log{p_{\theta}(\textbf{w}|\textbf{c})} \geq \mathbb{E}_{q_{\phi(\textbf{z}|\textbf{w})}}\left[\log{p_{\theta}(\textbf{w}|\textbf{z})}-\log{\frac{q_{\phi}(\textbf{z}|\textbf{w})}{p_\theta(\textbf{z}|\textbf{c})}}\right]
\end{equation}
In (5), the conditional information $\textbf{c}$ includes $[\textbf{c}_{text}, \textbf{A}]$, where $\textbf{c}_{text}$ represents the input text, and $\textbf{A}$ denotes the alignment between the input text and the target speech. The latent variable~$\textbf{z}$ is extracted from the posterior encoder during training and from the prior encoder during inference. The parameters~$\theta$ correspond to those of the prior encoder and the decoder, while $\phi$ indicates the parameters of the posterior encoder. The posterior encoder approximates the posterior of $\textbf{z}$ from the linear spectrogram of $\textbf{w}$. The prior encoder emulates $q_{\phi}(\textbf{z}|\textbf{w})$, using a text encoder, a normalizing flow, and a duration predictor. The alignment $\textbf{A}$ is either estimated by the monotonic alignment search~(MAS) algorithm during training or inferred by the duration predictor during inference. The normalizing flow of the prior encoder enhances the expressiveness of the prior encoder. The decoder generates the output waveform from the latent variable~$\textbf{z}$.

\begin{figure}[]
 \centering
 \includegraphics[width=\columnwidth]{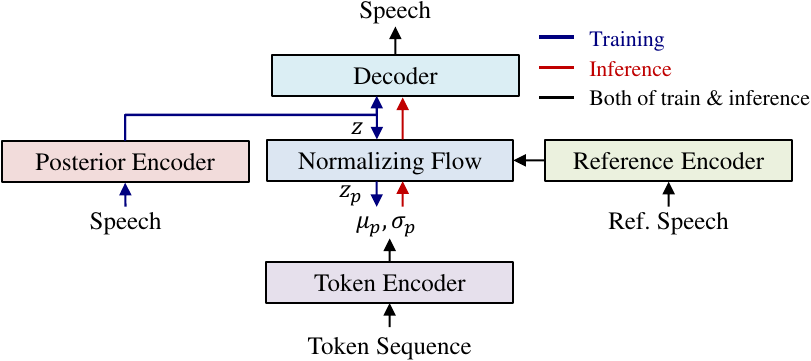}
 \caption{Architecture of the speech generator.}
 \label{fig:t-SNE}
 \end{figure}

\subsubsection{\bf{Architecture}}
We leverage the architecture of VITS for our speech generator with several modifications, as depicted in Fig.~3. First, we replace the input text~$\textbf{c}_{text}$ with the semantic token sequence $\textbf{c}_{token}$ and, correspondingly, substitute the text encoder with a token encoder. The token encoder shares the same architecture as the posterior encoder. In addition, we do not use the MAS algorithm and a duration predictor for alignment issues. Another alteration involves the use of reference speech to control acoustic conditions. We employ a reference encoder to extract the reference embedding, which adheres to the same architecture as the token transducer. The extracted reference embedding conditions the affine coupling layers of the normalizing flow, as in the original VITS. Thus, the conditional information is adjusted to $\textbf{c}=[\textbf{c}_{token}, \textbf{s}_{ref}]$ for the speech generator. It is worth noting that although both reference encoders in the token transducer and the speech generator share the same architecture, they extract different attributes of the reference speech for each objective.

\section{Experimental Settings}
\label{sec:majhead}

\subsection{Dataset}
We conducted zero-shot adaptive TTS experiments using the LibriTTS~\cite{zen2019libritts} corpus. LibriTTS is a multi-speaker English speech corpus with a 24 kHz sample rate. The dataset is partitioned into subsets of \textit{clean} and \textit{other}, depending on the recording conditions. Both subsets were employed to assess scalability and robustness across various recording conditions. For training, we utilized all of the training subsets, including train-clean-100, train-clean-360, and train-other-500, which amounts to a total of 555 hours of recordings spoken by 2,311 speakers. For evaluation, we used the test-clean and test-other subsets, which contains approximately 15 hours of speech data from 72 speakers, ensuring no overlap with speakers in the training set. All scripts were converted to phoneme sequences in the International Phonetic Alphabet (IPA) using the phonemizer library\footnote{\url{https://github.com/bootphon/phonemizer}}.

\subsection{Implementation Details}
\subsubsection{\bf{Semantic token}}
We obtained semantic tokens by leveraging the pre-trained wav2vec2.0-XLSR~\cite{conneau2020unsupervised}\footnote{\url{https://huggingface.co/facebook/wav2vec2-xlsr-53-espeak-cv-ft}} model. We applied $k$-means clustering to the representations extracted from the 15th block using the Faiss library\footnote{\url{https://github.com/facebookresearch/faiss}} and designated the cluster index as the semantic token. The choice of the 15th block was based on \cite{baevski2021unsupervised} due to its superior phoneme classification accuracy. We configured the number of clusters to be $k=512$ for our experiments.

\subsubsection{\bf{Token transducer}}
We implemented the neural transducer using the icefall toolkit\footnote{\url{https://github.com/k2-fsa/icefall}}. For the text encoder, we adopted a modified version of the conformer as the icefall implementation. Since this modified conformer is not officially documented, we encourage interested readers to refer to the source code\footnote{\url{https://github.com/k2-fsa/icefall/blob/master/egs/librispeech/ASR/pruned_transducer_stateless6/conformer.py}} for a more detail. The text encoder comprised 6 conformer blocks with 384 hidden dimensions, 1536 feed-forward dimensions, and a kernel size of 5. The prediction network, implemented in two different structures, had either 2 layers of unidirectional LSTM with 512 hidden dimensions for the LSTM version or a conformer configuration identical to the text encoder but with causal masking for the conformer version. The joint network consisted of 3 feed-forward blocks with 512 hidden dimensions. In the context of pruned training, we found that a pruning constant of $S=50$ was sufficient for our experimental setup. We trained the token transducer for 30 epochs with a dynamic batch size, containing up to 240 seconds of speech per iteration. We used the Adam optimizer with $\beta_1=0.9$, $\beta_2=0.98$, and a learning rate of 0.05. Training each model with LSTM or conformer prediction network took approximately 2-3 days on 2 Quadro RTX 8000 GPUs. During the training process, the target speech, randomly cropped to 3 seconds, was used as the reference speech $\textbf{s}_{ref}$. The cropping strategy mitigated the risk of target information leakage from the reference speech during training, as described in Section V-E. During inference, we employed top-$k$ sampling~\cite{fan2018hierarchical} with $k=5$ for autoregressive (AR) generation.

\subsubsection{\bf{Speech generator}}
We followed the configuration of the original VITS~\cite{kim2021conditional} with several modifications. We configured the intermediate features to have a frame length of 10 ms. Since semantic tokens share the frame length of 20 ms, consistent with wav2vec2.0, we performed a double upsampling of the semantic tokens using transposed convolution to match the frame length. Furthermore, considering the sample rate of 24~kHz, we adjusted the upsampling rates of the decoder to $(10, 6, 2, 2)$. We trained the speech generator for 400~k iterations with a batch size of 64, which took approximately 6 days on 2 Quadro RTX 8000 GPUs.

\subsection{Baselines}
We built two baseline models for comparison: VITS and VALL-E~\cite{wang2023neural}. To adapt the baseline VITS to the zero-shot adaptive scenario, we incorporated the ECAPA-TDNN structure as a reference encoder. Consequently, the architecture of the baseline VITS closely resembled that of our speech generator, except for the text encoder, duration predictor, and the alignment based on the monotonic alignment search~(MAS) algorithm. We used this baseline VITS to evaluate the effectiveness of using semantic tokens generated by our token transducer compared to using text input directly.

For the baseline VALL-E, we employed an unofficial implementation\footnote{\url{https://github.com/lifeiteng/vall-e}}. VALL-E represents a state-of-the-art TTS model for zero-shot speaker adaptation using in-context learning~\cite{wang2023neural}. It employs a hierarchically structured generation process, combining AR and NAR phases, with alignment modeling conducted during the AR phase. The baseline VALL-E was trained on the same LibriTTS dataset over approximately 4 days, utilizing 8~Tesla A100 GPUs.

\section{Results}

\subsection{Main Results: Overall Performance}
We evaluated the performance of the proposed model based on both subjective and objective measures. For subjective measures, we conducted a mean opinion score~(MOS) test and a similarity mean opinion score~(SMOS) test. In the MOS test, 17 listeners rated the quality of 30 randomly selected samples per model on a 5-point scale regarding naturalness and intelligibility. In the SMOS test, the listeners evaluated the similarity of speaker and prosody between the reference and generated samples on the same 5-point scale. For objective measures, we calculated character error rate~(CER) and speaker embedding cosine similarity~(SECS). CER is an indicator of sample intelligibility by an ASR model, with the samples being transcribed by the whisper-large model~\cite{radford2022robust}. Meanwhile, SECS was measured by a speaker verification model from speechbrain~\cite{ravanelli2021speechbrain}. SECS values range from -1 to 1, with a higher score indicating higher similarity. The results are shown in Table~I. The proposed models are labeled as ``proposed-lstm" and ``proposed-conformer," signifying the version of the prediction network of the token transducer. Unless specified otherwise, we refer to ``proposed-lstm" as the proposed model 
from now on.

\begin{table}[h]
\setlength{\tabcolsep}{5pt}
\setlength{\arrayrulewidth}{0.2mm}
\caption{Results of zero-shot adaptive TTS. MOS and SMOS are represented with 95\% confidence intervals.}
\label{tab:Model size}
\centering
\begin{tabular}{l c c c c}
\toprule
\textbf{Method} & \textbf{MOS}& \textbf{SMOS} & \textbf{CER(\%)}& \textbf{SECS} \\ 
\midrule
Ground Truth & 4.35\footnotesize{$\pm$0.07} & 4.36\footnotesize{$\pm$0.08} & 1.45 & 0.653   \\
\midrule
VITS      &3.59\footnotesize{$\pm$0.07} & 4.07\footnotesize{$\pm$0.08} & 5.41 & 0.491                             \\
VALL-E     &3.18\footnotesize{$\pm$0.09} & 3.42\footnotesize{$\pm$0.09}  & 25.54 & 0.347   \\
\midrule
Proposed-lstm     &\textbf{4.24\footnotesize{$\pm$0.06}} &4.31\footnotesize{$\pm$0.07}   & \textbf{3.23} & \textbf{0.498}      \\
Proposed-conformer     &4.19\footnotesize{$\pm$0.06} &\textbf{4.34\footnotesize{$\pm$0.07}}   & 3.29 & 0.494      \\
\bottomrule
\end{tabular}
\end{table}

As presented in Table~I, the proposed models consistently outperformed the baselines across all the metrics related to sample quality and speaker similarity. Two versions of the proposed model demonstrated similar performance in all the metrics with only marginal differences. This trend suggests that the complexity of the prediction network had a limited influence on the overall performance. It is likely that the simplified information contained within the semantic tokens, compared to waveform, places greater importance on the text encoder and the joint network within the token transducer. When compared to the VITS baseline, the performance improvements underscore the effectiveness of leveraging semantic tokens generated by the token transducer. In contrast, VALL-E exhibited the lowest sample quality and occasionally generated misaligned samples, particularly with word skipping, significantly impacting its overall performance. This tendency is attributed to the absence of strict monotonic alignment constraints within VALL-E. It is worth noting that the baseline VALL-E showed a performance degradation compared to the results reported in the original paper, due to the unofficial implementation being trained on a substantially smaller dataset with fewer computational resources.

\subsection{Analysis: Alignment of Token Transducer}
We visualized examples of alignments produced by the token transducer in Fig.~4. Fig.~4(a) and Fig.~4(b) demonstrate the monotonic alignment tendencies between the given text and semantic tokens. The node gradient in Fig.~4(b) reveals the critical nodes among the marginalized paths, which can be interpreted as an alignment visualization as in \cite{kuang2022pruned}. Fig.~4(a) and Fig.~4(b) show smoothed monotonic regions and continuous values, considering marginalized paths. In contrast, the generated alignment in Fig.~4(c) has binary values because the inference process propagates toward a single path by sampled transitions and emissions.

\begin{figure}[h]
 \centering
 \includegraphics[width=\columnwidth]{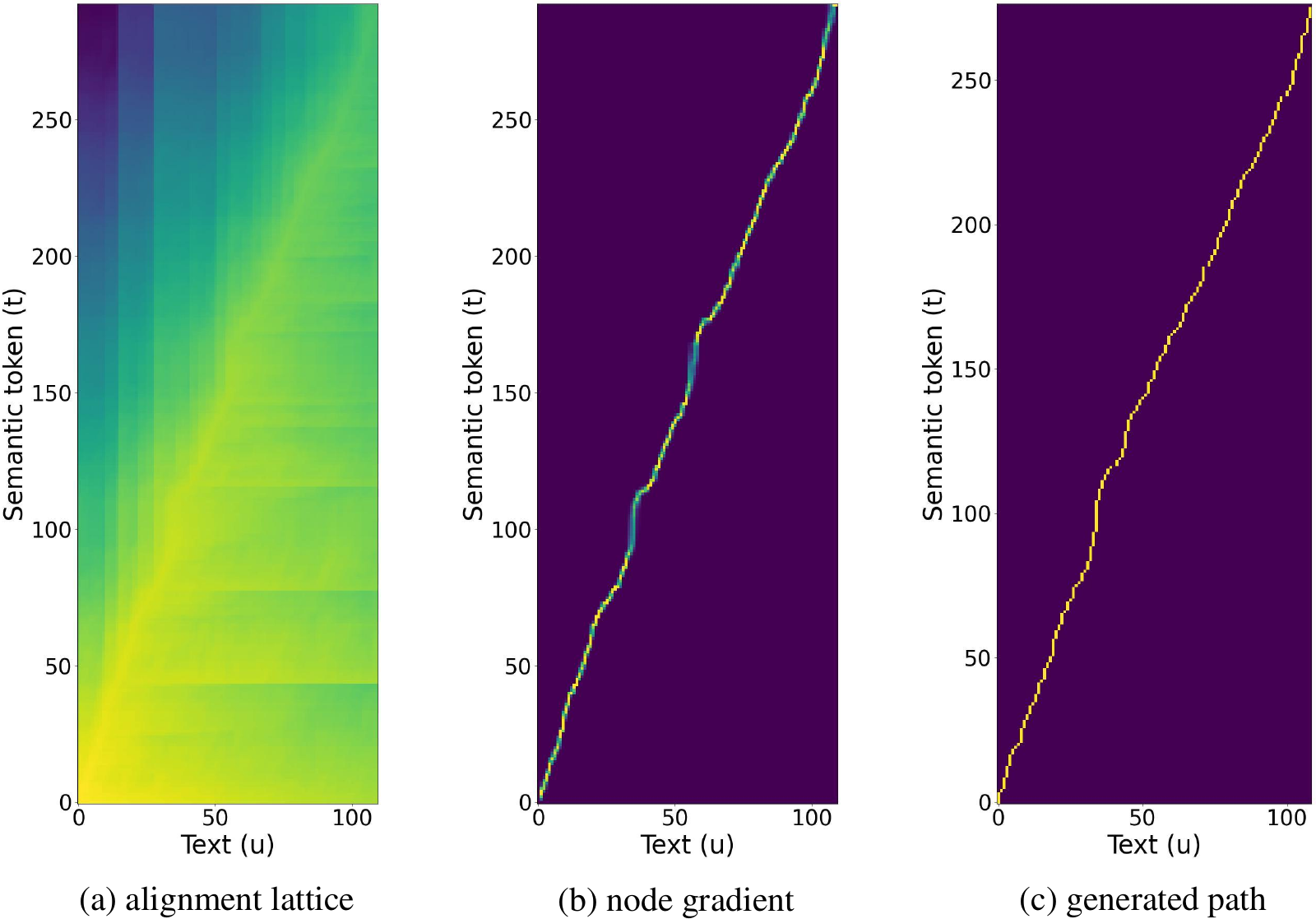}
 \caption{Examples of alignment produced by the token transducer. (a) Alignment lattice of $\log\alpha(u,t)$ between text and semantic tokens. (b) Node gradient of negative log likelihood. (c) Generated alignment.}
 \label{fig:t-SNE}
 \end{figure}

To assess the alignment's robustness of the generated samples, we analyzed the elements of word-level Levenshtein distance~\cite{levenshtein1966binary} between the ground truth text and transcribed text by the whisper-large model in Table~II. We focused on insertion~(INS) and deletion~(DEL) errors of the Levenshtein distance because INS and DEL errors primarily result from mismatched word counts due to misalignment. INS errors can occur when the TTS model generates unnecessary words, as if repeating, and DEL errors can occur when the transcription is much shorter than the ground truth due to word skipping and early termination. Substitution (SUB) errors mainly result from mispronunciation and low intelligibility. In Table~II, VITS and the proposed models exhibited similarly small INS and DEL errors, indicating robust alignment. In contrast, VALL-E exhibited numerous alignment errors, especially word skipping and early termination, due to its unconstrained AR method. We acknowledge that Levenshtein distance has limitations when measuring alignment robustness, since other factors can contribute to the Levenshtein distance, such as transcription errors. Nevertheless, in our considered view, the elements of Levenshtein distance provide valuable insights into overall alignment issues.

\begin{table}[h]
\setlength{\tabcolsep}{5pt}
\setlength{\arrayrulewidth}{0.2mm}
\caption{Comparison of elements of Word-Level Levenshtein distance and Word Error Rate (WER).}
\label{tab:Model size}
\centering
\begin{tabular}{l c c c c}
\toprule
\textbf{Method} & \textbf{INS (\%)}& \textbf{DEL (\%)} & \textbf{SUB (\%)}& \textbf{WER (\%)} \\ 
\midrule
Ground Truth & 0.49 & 0.69 & 2.10 & 3.29   \\
\midrule
VITS      &1.45 & 1.36 & 7.29 & 10.11                             \\
VALL-E     &2.05 & 18.43 & 10.43 & 31.22   \\
\midrule
Proposed-lstm     &0.76 & 1.17 & 4.66 & 6.59      \\
Proposed-conformer      &0.78 & 1.16 & 4.69 & 6.62      \\
\bottomrule
\end{tabular}
\end{table}

\subsection{Analysis: Inference Speed}
We compared the inference speed of the models, as presented in Table~III and Fig.~5. High computational requirements and lengthy inference times can limit the practical applications of even high-performance models, particularly for real-time or streaming services. To measure inference speed, we used a 5-second sample as the reference speech and calculated it for 1000 sentences from the test set. VITS exhibited the fastest inference speed by a significant margin due to its fully NAR architecture and the smallest model size. On the other hand, VALL-E had the slowest inference speed because it included a large-scale AR phase. The proposed models exhibited notable differences in inference speed depending on the version of the prediction network. The proposed-lstm was capable of generating speech much faster than the proposed-conformer, as LSTM requires substantially less computation for sequential prediction compared to the conformer, which has a complexity of $O(T^2)$. It is worth noting that a significant portion of the inference time was attributed to the token transducer. Despite being slower than the baseline VITS, the proposed-lstm was able to generate speech 10 times faster than real-time.

\begin{table}[h]
\setlength{\tabcolsep}{8pt}
\setlength{\arrayrulewidth}{0.2mm}
\caption{Comparison of average inference speed. All inference was conducted on a Quadro RTX8000 GPU. Speed (kHz) represents the number of raw audio samples per second, and ``real-time" indicates the inference speed relative to real-time. ``w/o speech generator" indicates the portion of inference without the token transducer.}
\label{tab:Model size}
\centering
\begin{tabular}{l c c c c}
\toprule
\textbf{Method} & \textbf{Speed (kHz)} & \textbf{Real-time}\\ 
\midrule
VITS        & 2311.04 & $\times$96.29              \\
VALL-E       & 7.53 & $\times$0.31              \\
\midrule
Proposed-lstm       & 246.60 & $\times$10.27             \\
\hspace*{0.2cm}w/o speech generator       & 273.84 & $\times$11.41             \\
\midrule
Proposed-conformer        & 24.99 & $\times$1.04              \\
\hspace*{0.2cm}w/o speech generator        & 25.26 & $\times$1.05              \\
\bottomrule
\end{tabular}
\end{table}

\begin{figure}[h]
 \centering
 \includegraphics[width=0.9\columnwidth]{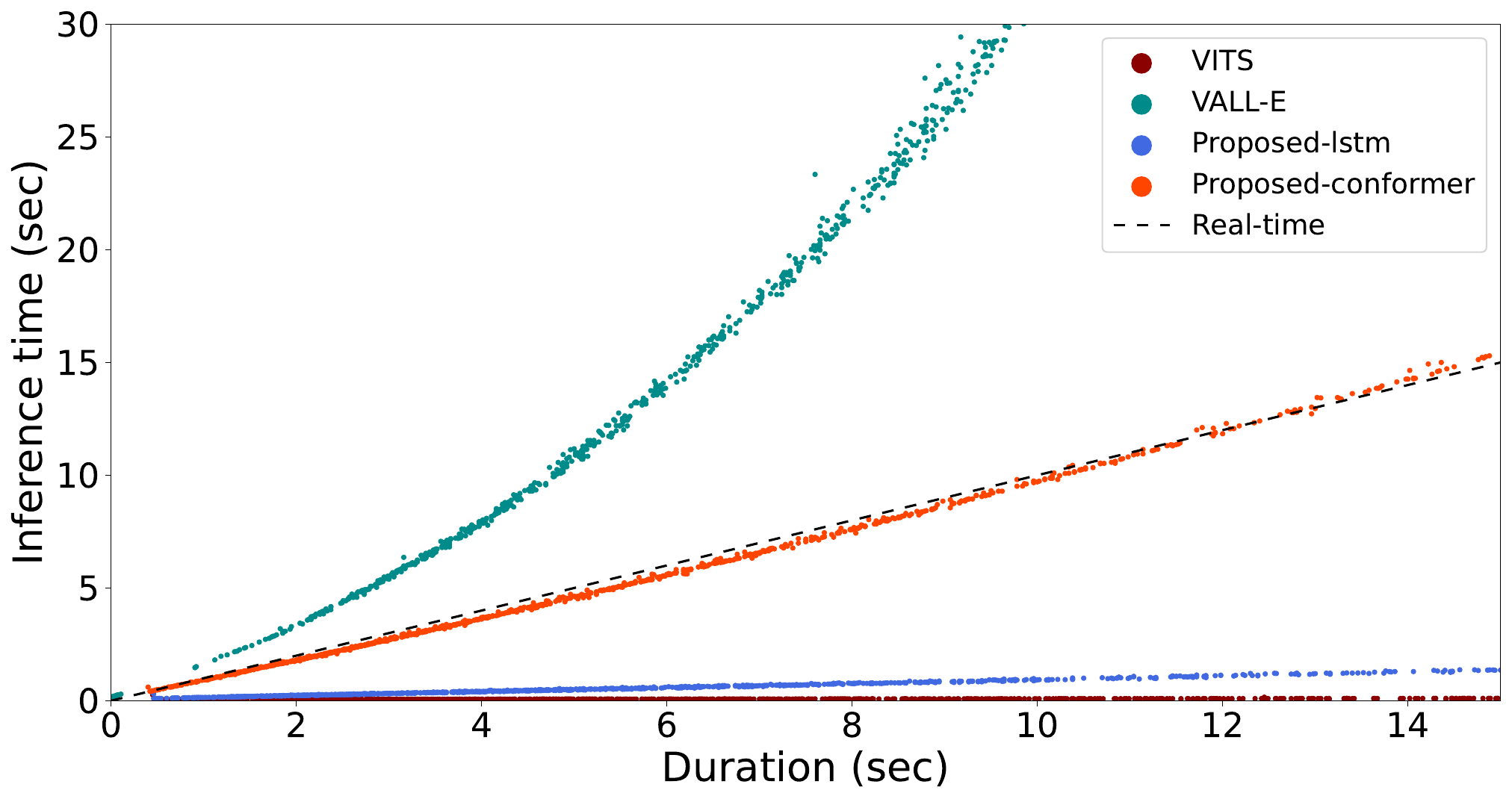}
 \caption{Visualization of inference speed. Each sample point represents the correlation between sample duration and inference time.}
 \label{fig:t-SNE}
 \end{figure}

\subsection{Analysis: Paralinguistic Controllability}
We investigated the controllability of the proposed framework as a zero-shot adaptive TTS model. The framework utilizes two reference speech: one for the token transducer and one for the speech generator, denoted as $\textbf{s}_{ref}^{tran}$ and $\textbf{s}_{ref}^{gen}$, respectively. In the basic case, the same reference speech is used for both components, meaning $\textbf{s}_{ref}^{tran} = \textbf{s}_{ref}^{gen}$. However, it is also possible to control the paralinguistic information in each component separately by using different $\textbf{s}_{ref}^{tran}$ and $\textbf{s}_{ref}^{gen}$. We demonstrated the controllability of each component in Fig.~6. As shown in Fig.~6(a) and Fig.~6(c), prosody-related temporal dynamics such as speech rate, inter-segment silence, and pitch contour are mainly influenced by $\textbf{s}_{ref}^{tran}$. In contrast, Fig.~6(b) and Fig.~6(d) show that samples generated from the same semantic token share the same alignment and similar pitch contour dynamics even when different $\textbf{s}_{ref}^{gen}$ is used. We also measured the SECS between the generated samples $\textbf{s}_{gen}$ and reference speech, as shown in Table~IV. The results in Table~IV indicate that speaker similarity is mainly influenced by $\textbf{s}_{ref}^{gen}$, while $\textbf{s}_{ref}^{tran}$ rarely shares speaker similarity with $\textbf{s}_{gen}$ and exhibited comparable scores to arbitrary samples $\textbf{s}_{rand}$. However, when $\textbf{s}_{ref}^{tran} \neq \textbf{s}_{ref}^{gen}$, the $\textbf{s}_{ref}^{gen}$ resulted in a slight degradation of SECS compared to the case when $\textbf{s}_{ref}^{tran} = \textbf{s}_{ref}^{gen}$. We conjecture this is because the temporal dynamics reflected from $\textbf{s}_{ref}^{tran}$ partially affect speaker similarity.

These results confirm that the temporal characteristics of the samples are mainly controlled by the token transducer, while global characteristics such as speaker information are controlled by the speech generator. We also observed the prosody controllability of the proposed token transducer. Additional auditory materials are available on our demo page to provide more detailed results.

\begin{figure*}[t]
\centering
\includegraphics[width=1.9\columnwidth]{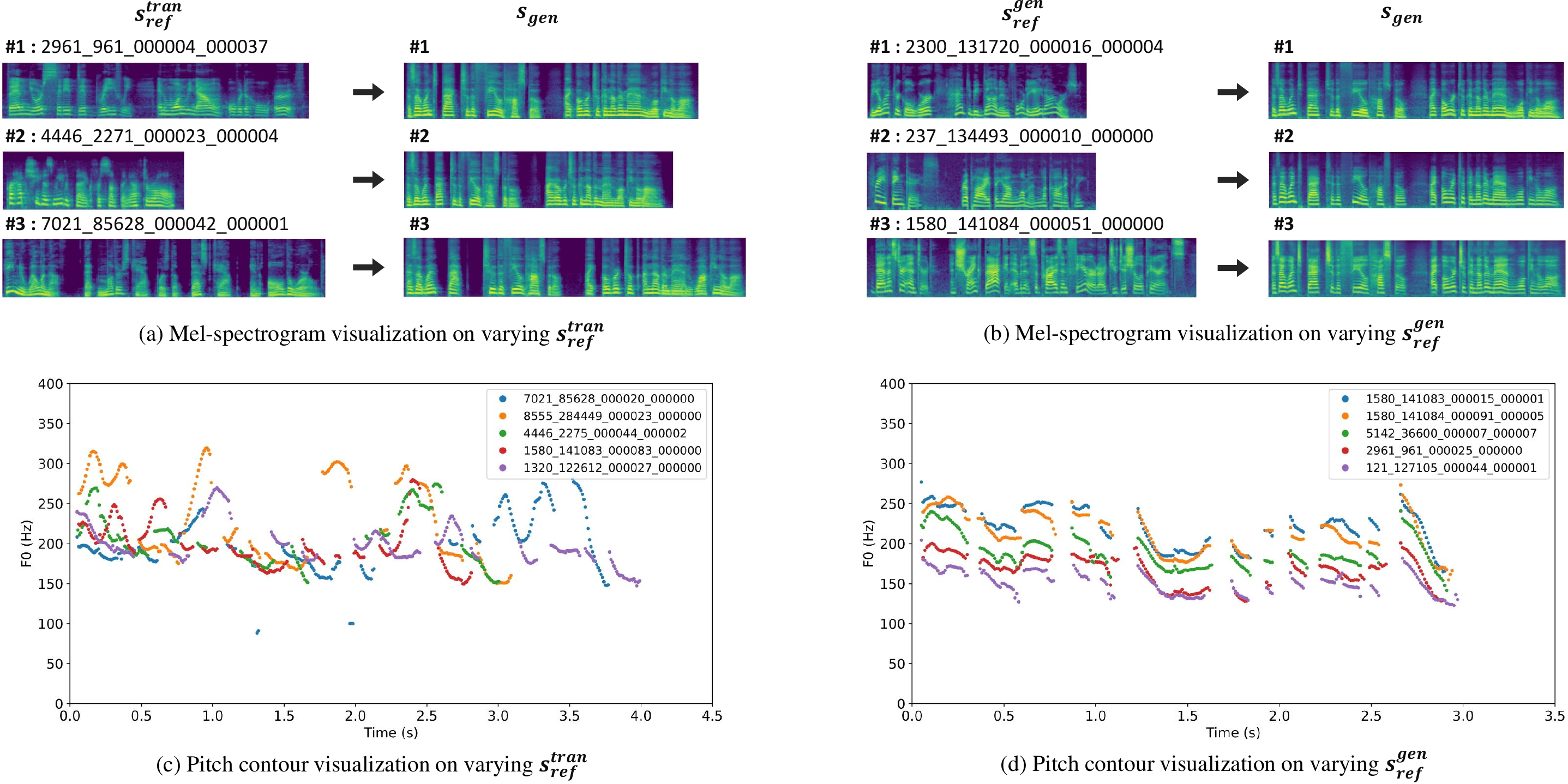}
\caption{Speech samples generated from varying reference speech. (a) and (b) show the generated mel-spectrograms of the utterance: ``As I went back to the field hospital, I overtook another man walking along.''. (c) and (d) are pitch contours of the utterance: ``You will have a very pleasant time, I am sure; because you never quarrel.''. In (a) and (c), the generated samples $\textbf{s}_{gen}$ share the same $\textbf{s}_{ref}^{gen}$ but are generated from varying $\textbf{s}_{ref}^{tran}$. In contrast, samples of (b) and (d) are extracted from the same semantic token sequence but varying $\textbf{s}_{ref}^{tran}$. The used $\textbf{s}_{ref}^{tran}$ and $\textbf{s}_{ref}^{gen}$ are denoted by utterance id of LibriTTS corpus.}
\label{zs_vc}
\end{figure*}

\begin{table}[h]
\centering
\caption{Comparison of speaker embedding cosine similarity~(SECS) between generated samples and the used reference speech. We randomly made 1000 tuples of (text, $\textbf{s}_{ref}^{tran}$, $\textbf{s}_{ref}^{gen}$) from the test set to generate utterances. $\textbf{s}_{rand}$ means arbitrary samples from the test set, and $\textbf{s}_{ref}^{tran}=\textbf{s}_{ref}^{gen}$ means the case of using same reference speech, both of which are used as standards.}
\label{tab:my-table}
\begin{tabular}{@{}cccll@{}}
\toprule
\multicolumn{2}{c}{{\textbf{Reference}}} & \multicolumn{3}{c}{{\textbf{SECS from $\textbf{s}_{gen}$}}} \\
\midrule
\multicolumn{2}{c}{$\textbf{s}_{rand}$}                              & \multicolumn{3}{c}{0.078}                          \\ \midrule
\multicolumn{2}{c}{$\textbf{s}_{ref}^{tran}=\textbf{s}_{ref}^{gen}$}                              & \multicolumn{3}{c}{0.498}                          \\ \midrule
\multirow{2}{*}{$\textbf{s}_{ref}^{tran}\neq\textbf{s}_{ref}^{gen}$} & $\textbf{s}_{ref}^{tran}$ & \multicolumn{3}{c}{0.095}
\\ \cmidrule(l){2-5} & $\textbf{s}_{ref}^{gen}$ & \multicolumn{3}{c}{0.454}
\\ \bottomrule
\end{tabular}%
\end{table}

\begin{table}[h]
\centering
\caption{Comparison of Character Error Rate~(CER) and Negative Log Likelihood~(NLL) of target semantic tokens based on the reference speech. ``Cropping" indicates whether the target speech was randomly cropped to 3 seconds for $\textbf{s}_{ref}$ during training. ``Text matched" signifies that the reference speech shares the same script as the target script for inference, whereas the "text mismatched" case represents the general scenario for zero-shot adaptive TTS.}
\label{tab:my-table}
\begin{tabular}{@{}cccllc@{}}
\toprule
\multicolumn{2}{c}{{\textbf{Reference}}} & \multicolumn{3}{c}{{\textbf{CER (\%)}}} & {\textbf{NLL}} \\
\midrule
\multirow{2}{*}{Trained without cropping}  & Text matched & \multicolumn{3}{c}{2.64}  & 1.14                   \\
\cmidrule(l){2-6}  & Text mismatched & \multicolumn{3}{c}{9.02}  & 2.75                   \\
\midrule
\multirow{2}{*}{Trained with cropping (proposed)} & Text matched & \multicolumn{3}{c}{2.77} & 1.22  \\
\cmidrule(l){2-6} & Text mismatched & \multicolumn{3}{c}{3.53}   & 1.50 \\
\bottomrule
\end{tabular}%
\end{table}

\subsection{Ablation: Cropped Reference Speech for Token Transducer}
The token transducer utilizes a reference speech, denoted as $\textbf{s}_{ref}$, to control prosodic information. During training, we employed the target speech as $\textbf{s}_{ref}$ because it ensures prosodic similarity between $\textbf{s}_{ref}$ and the target semantic tokens. However, we discovered that linguistic information from the target semantic tokens leaked from $\textbf{s}_{ref}$ within our framework. This leakage occurred because $\textbf{s}_{ref}$ shared linguistic information with the target semantic tokens. This issue led to inaccurate pronunciation during inference as the token transducer learned to extract phonetic information from $\textbf{s}_{ref}$, which is not accessible during inference. As a remedy, we randomly cropped the target speech to be 3 seconds during training to interrupt the leakage while partially maintaining prosodic similarity.  To quantify the leakage of linguistic information, we measured Character Error Rate (CER) of the generated samples and negative log likelihood (NLL) of target semantic tokens given text and $\textbf{s}_{ref}$ as indicated in $\mathcal{L}_{tran'}$ for several reference speech scenarios, as shown in Table~V. The token transducer trained without cropping exhibited superior performance in the text-matched case, but significantly deteriorated when text was mismatched to $\textbf{s}_{ref}$. This trend highlights the issue of linguistic information leakage. In contrast, the token transducer trained with cropping showed a much smaller performance gap and higher accuracy in the text-mismatched case. We emphasize that this information leakage is not a specific phenomenon of the neural transducer but is a general concern in zero-shot adaptive TTS. This issue can be resolved through alternative approaches such as different reference conditioning or dataset composition. In our work, we proposed the technique of cropping the reference speech for the token transducer.

\section{Discussion and Conclusion}
\label{sec:ref}
We proposed a novel TTS framework that uses a neural transducer to predict semantic tokens. The proposed framework consists of a token transducer and a speech generator, with semantic tokens positioned in between. The token transducer primarily provides alignment modeling and temporal prosody control, while the speech generator is responsible for generating waveforms and controlling acoustic conditions, including speaker identity. We verified the proposed model in terms of speech quality, inference speed, and the controllability of each component. From the experimental results, we found that the LSTM-based prediction network showed the best performance with rapid inference speed.

The two parts of the proposed framework form a cascading system, allowing us to enhance each component separately. In our future work, we plan to improve the speech generator by exploring alternative structures using acoustic tokens from neural codecs. Additionally, we are interested in refining the token transducer by investigating more sophisticated schemes to achieve greater prosody controllability.

\section*{Acknowledgments}
This work is/was supported by SAMSUNG Research, Samsung Electronics Co.,Ltd.

\bibliographystyle{unsrt}
\bibliography{refs}

\end{document}